# Heavy fermion material: Ce versus Yb case

J Flouquet [1], H Harima [2]

[1] *INAC/SPSMS, CEA-Grenoble, 17 rue des Martyrs, 38054 Grenoble, France*
[2] *Department of Physics, Graduate School of Science, Kobe University, Kobe, Hyogo 657-8501, Japan*

Abstract: Heavy fermion compounds are complex systems but excellent materials to study quantum criticality with the switch of different ground states. Here a special attention is given on the interplay between magnetic and valence instabilities which can be crossed or approached by tuning the system by pressure or magnetic field. By contrast to conventional rare earth magnetism or classical s-wave superconductivity, strong couplings may occur with drastic changes in spin or charge dynamics. Measurements on Ce materials give already a sound basis with clear key factors. They have pointed out that close to a magnetic or a valence criticality unexpected phenomena such as unconventional superconductivity, non Fermi liquid behaviour and the possibility of re-entrance phenomena under magnetic field. Recent progresses in the growth of Yb heavy fermion compounds give the perspectives of clear interplays between valence and magnetic fluctuations and also the possibility to enter in new situations such as valence transitions inside a sole crystal field doublet ground state.

## 1/ Introduction

The interest in heavy fermion materials started three decades ago with the discovery that in a compound ($CeAl_3$) (1) (figure 1) the extrapolation of the Sommerfeld coefficient γ of the ratio S/T of the entropy (S) by the temperature (T) reaches a value γ near 1 J·mole$^{-1}$K$^{-2}$ i.e. three orders of magnitude higher than that of noble metals like Cu. As γ ∝ $m^*$, the mass of the itinerant quasiparticle, that suggests heavy quasiparticles moving with a huge effective mass ($m^* = 10^3 \, m_o$, $m_o$ being the free electron mass). During few years, some physicists argued about whether the strength of the γ term can be related to the effective mass $m^*$ of slow quasiparticles. From the observation of quantum oscillations in so-called de Haas van Alphen experiments (dHvA), highly used previously to study the properties of the Fermi surface (both, the topology and effective mass measurements), it was established that the previously proposed image works: heavy quasiparticle moves on well defined orbits of the Fermi surface (2,3). An unexpected new perspective emerges a few years later with the discovery of the superconductivity (SC) in $CeCu_2Si_2$ (4) (figure 2) implying clearly the Cooper pairing of the heavy particles (5). This observation opened the field of unconventional super-conductivity which covers now also the large domain of high $T_C$ superconductors (6), organic conductors (7) and maybe the new recently discovered materials of pnictide superconductors (8). It is an extension of the class of non s-wave pairing of superfluid $^3$He with its triplet pairing and opens also the new window to study the charge motion, a phenomenon absent in the case of the $^3$He neutral atom.

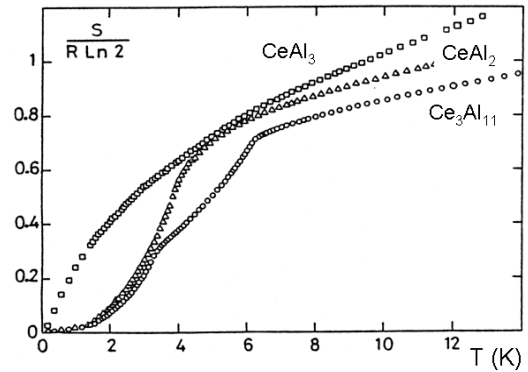

Fig. 1 : Temperature variation of the entropy S normalized to the maximum value R ln2 given by the 2S+1 degree of freedom of a doublet of $S = \frac{1}{2}$. For the $CeAl_3$, S decreases smoothly on cooling with huge linear γT dependence at very low temperature; for $CeAl_2$ and $Ce_3Al_{11}$, two magnetically ordered compounds, a drop of entropy is observed at the onset of each magnetic phase transition (9).

A key ingredient for this heavy fermion creation is the transition from a single impurity behaviour (see later the Kondo effect with the appearance of the Kondo temperature $T_K$) to a regular array of rare earth ions like Ce or Yb where the unfilled occupation of the 4f shell (14 electrons for a complete occupancy) is a main factor for the development of long range magnetism. When the occupation number of the 4f shell is integer, intermetallic rare earth compounds present their classical behaviour. These are still highly studied for their magnetic properties. Anomalous rare earth behaviours with the formation of heavy quasi particles have been reported mainly for Ce and Yb ions; there are also examples for Pr, Sm and Tm.

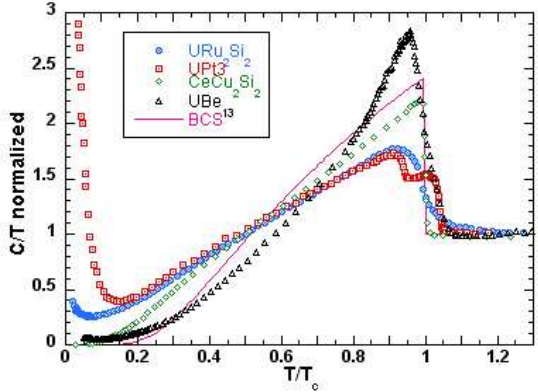

Fig. 2: Temperature variation of the specific heat C divided by the temperature for four heavy fermion superconductors: $CeCu_2Si_2$, $UBe_{13}$, $UPt_3$, and $URu_2Si_2$. By comparison to the so called BCS behaviour for s wave superconductors with a full gap opening in all directions of the wavevectors, $C/T$ does not present an exponential decrease (as described by the s wave BCS curve) but a power law dependence with $T^n$ which is directly related to the presence of line nodes or point nodes in these anisotropic superconductors (9).

We restrict the discussion to Ce and Yb heavy fermion compounds (HFC) where the Ce or Yb centers are close to their trivalent configuration $Ce^{3+}$ or $Yb^{3+}$. A key parameter is the occupation number $n_f$ of the trivalent configuration. For $n_f$ near unity (large $n_f$), the name of Kondo lattice is given to a Ce or Yb lattice by extension to the Kondo problem of a single trivalent impurity dissolved in the Fermi sea of a non magnetic host (for example Ce in $LaAl_3$). The domain of heavy fermion compounds covers also the cases of uranium and even transuranian elements with their incomplete 5f shells (Np, Pu). The discovery of unconventional superconductivity two decades ago in three uranium HFC: $UBe_{13}$, $UPt_3$, $URu_2Si_2$ (see 9) (figure 2) plays a major role in the establishment and developments of exotic pairing thanks to the possibility to grow excellent large crystals. It is also in the U series that ferromagnetic superconductors have been discovered ($UGe_2$, URhGe, UCoGe). Even it is on "hot radioactive" materials like $PuCoGa_5$ (10) or $NpPd_5Al_2$ (11) that the highest critical superconducting temperatures $T_C$ were found.

Here we have restricted the discussion to few examples of Ce and Yb HFC with the aim to show why the research of the ideal material will open new roads. It is of course a good domain for an optimist: tomorrow will be better… The selection of materials is directly connected with the activity in Grenoble and often with the feedback with Kobe-Osaka colleagues. All the examples correspond to materials with high symmetry: tetragonal ($CeRu_2Si_2$, 115 Ce compounds, $YbAlB_4$, $YbCu_2Si_2$, $YbRh_2Si_2$) or cubic (Ce metal, $YbXCu_4$ series). References to $CeRu_2Si_2$ are due to the fact that it was extensively studied in Grenoble but also in Osaka, Tokyo, mostly on competing basis and sometimes in collaboration. Work on $YbCu_2Si_2$ have been made recently in Grenoble, however, the first input comes from measurements realized in Cologne and Geneva; it has been revisited in Osaka- Kobe quite recently. At least 90% of the experimental publications on $YbRh_2Si_2$ have been made by letters from results obtained in Dresden with strong theoretical impacts from Rutgers, Houston and Saclay; in Grenoble, our choice was to present an unique article which quite different views that the previous ones. The discovery of SC in $YbAlB_4$ was recently made in Tokyo through a large international network (Irvine, Cambridge, Kyoto). Our interest on Ce metal phase diagram coincides with the attempt to draw the main road of heavy fermion physics in a review article (9). Interest in $YbInCu_4$ (highly studied in Los Alamos and Kobe) started with the idea that the magnetic field itself may influence the valence transition. This idea has been recently put in a theoretical frame by our theoretician friends in Osaka and Tokyo. A sound view was always to verify if our over simplification is supported by band structure calculations (strong collaboration with Kobe) and if dHvA data are available in agreement with Fermi Surface measurements mainly realized in Osaka.

## 2/ Complex matter but unique matter for quantum phase transitions.

Studies on heavy fermion matter enter in the wide domain of low temperature physics (9). Due to their low characteristic temperatures, the great interest of heavy fermion compounds is that their physical properties can change drastically under pressure ($P$) and also magnetic field ($H$). They are excellent materials to study ($T$, $P$, $H$) quantum phase transition (QPT) (12). We will use the abbreviation FOT and QCP respectively for first order and second order phase transition with the additional label of M or V when it will be driven by magnetic or valence fluctuation: M-FOT, M-QCP for magnetism, V-FOT and V-QCP for valence; the symbol CEP or TCP will correspond respectively to critical end point and tricritical point. Two main aspects are the determination of their ground state and the associated low energy excitations.

Due to the hybridization with the moving light electrons, the initially localized 4f electrons can escape from its deep potential and can move now very slowly (with the Fermi velocity $v_F$) with a heavy mass ($m^*$) ($m^* v_F = k_F$ the Fermi wavevector which is basically the inverse $a^{-1}$ of an atomic distance)(figure 3a). The large magnetic entropy ($S = R\ln2$) given by the degree of freedom of the spin

$S = ½$ of the bare localized 4f electrons instead of collapsing via a drop at a magnetic ordering temperature is transferred below a low temperature in kinetic energy such as if the Kondo effect is also efficient for a lattice . $T_K$ being defined by the temperature where the entropy reaches one half of the maximum value $R\ln 2$ of a spin 1/2, roughly $R\ln 2$ is comparable to $\gamma T_K$ and thus $\gamma = m^*$ may lead to a huge effective mass (figure 1) for low enough $T_K$. As we will see a part of the effective mass enhancement will be given by spin or valence fluctuations characteristic of magnetic or valence instabilities of the lattice.

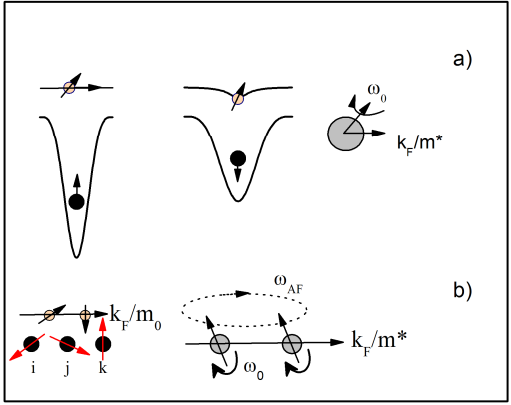

Fig. 3 : Scheme representing :
a) the formation of the heavy quasiparticles. Initially, the light conduction electron (o with its spin s) on approaching the localized 4f electron deforms the potential. That leads to a dressed heavy quasiparticle which moves very slowly while its angular momentum fluctuates at an energy $\omega_0 \sim k_B T_K$,
b) the transmission of the magnetic information is carried from site to site by the light electron. That leads to a coupling between the heavy particles with an additional precession at a frequency $\omega_{AF}$ which will collapse at the onset of long range magnetic ordering.

In HFC, the reasoning concerns mainly the sole 4f electron with the shortcut that it is not the only part of the initial problem which involves also s,p,d electrons . For example in addition to the Kondo local fluctuations, the fast moving light electrons will carry the magnetic interaction $E_{ij}$ via the so-called Ruderman Kittel Kasuya Yosida oscillations from site $i$ to site $j$. At the long life time of the heavy quasiparticle, they seem to interact themselves fluctuating locally at a frequency $\omega_0 = k_B T_K$ but becoming synchronized at a frequency $\omega_{AF}$ for the wavevector which is at the origin of large magnetic fluctuation (figure 3b). By reference to optics, we have the association of a strong local source of diffraction and interference between two sources.

Of course multi-band phenomena exist but at first approximation the other carriers are involved in the building of the many body effects such the Kondo effect ($T_K$), the valence transition ($T_v$) or the intersite magnetic coupling ($E_{ij}$). Outside HFC compounds, there are still active researches on classical 4f intermetallic compounds but they involve the situation of 4f electrons being fully localized (or $T_K \rightarrow 0K$). Their actions on itinerant electrons are limited to simple molecular interaction. The novelty of HFC is that the same quasiparticle (built by complex many body phenomena) carries the magnetism and the current so somewhere on cooling the bottleneck will arrive with a strong coupling behaviour as for the single Kondo impurity below $T_K$. A nice illustration of the difference between classical or anomalous 4f behaviour is the interplay between SC and magnetism. In conventional magnetic super-conductors like the Chevrel phases highly studied three decades ago, even if SC appears at $T_c$ above the establishment of AF at $T_N$, the restricted number of electrons involved in the pairing ($N(E_F) \Delta_{SC} < 1\%$, $\Delta_{SC}$ being the SC gap) lead to a low SC energy $N(E_F) \Delta_{SC}^2 \sim 10^{-2} k_B T_c$ per volume unit by comparison to the magnetic energy $\sim k_B T_N$ (100% of magnetic sites involved). This well-known peaceful life between antiferromagnetism and SC will be totally broken in the heavy fermion matter as the renormalized density of states involved all the electrons: $\Delta_{SC}/k_B T_K \sim 1$. So a strong interplay must exist. If there is coexistence, it may lead to a complete readjustment in spin dynamics and current circulation.

## 3/ Valence transition always present but often hidden

At first glance, we assume that the ground state of a HFC is a paramagnetic (PM) Fermi liquid: the electrons are ordered in wavevector according to Fermi statistics up to the Fermi wave vector. Of course a competing ground state can be a magnetic ordered state either antiferromagnetic (AF) or ferromagnetic (FM) (it is well-known that classical rare earth intermetallic systems play a major role in magnet technology). Another possibility is to modify the occupation number $n_f$ of the trivalent configuration via a so called valence transition $n_f$ changes discontinuously from one value to another. For Ce HFC, the stable configuration at large volume (low pressure) is the trivalent one, with a possible mixing with the tetravalent state at high pressure according to the relation:

$$Ce^{3+} \rightleftarrows Ce^{4+} + 5d$$

The valence is defined by $v = 4 - n_f$. For Yb HFC, $Yb^{2+}$ is the stable configuration at large volume while $Yb^{3+}$ will be achieved at high pressure according to the valence balance:

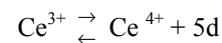

$$Yb^{2+} \underset{\leftarrow}{\rightarrow} Yb^{3+} + 5d$$

$v = 2 + n_f$, $n_f$ being now the hole numbering of the 4f shell. The asymmetry in the liberation or absorption of a 5d electron for a deviation of $n_f$ from unity between Ce and Yb is a main point in their pressure behaviour as described below.

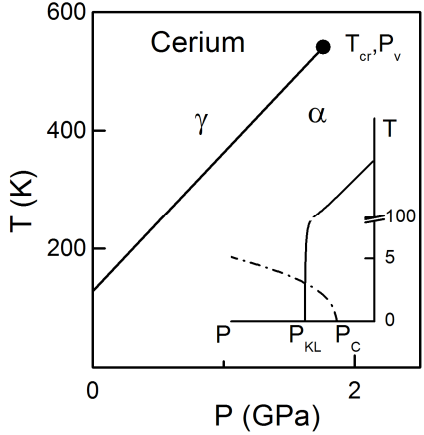

Fig. 4: $(T, P)$ phase diagram of Ce metal with the first order transition $T_V(P)$ between the γ and α phase and its critical end point. If the volume could be expanded at negative pressure: (i) $T_V(P)$ would collapse abruptly to zero at $T \to 0$ K at $P = P_{KL}$ and (ii) magnetic ordering must occur in the γ phase up to a pressure $P_C$ (9).

De facto one can consider that the class of heavy fermion materials belongs to the general class of the anomalous 4f lattice which was recognized six decades ago via the discovery of the Ce metal phase diagram (see 13-14) schematically shown in figure 4. The first order transition at $T_V(P)$ from a PM γ phase to a PM α phase is isostructural; on warming it ends up at a critical end point (CEP) near $T_{cr} = 600$ K, $P_V \sim 2$ GPa. From high energy spectroscopy it is now well established that in the γ state $n_f$ is near unity while even in the α phase (referred as small $n_f$) $n_f$ remains rather high ($n_f \sim 0.8$) (15). As in γ phase, $n_f$ is near unity, one may expect that if the intercept $T_{γα}$ at $P = 0$ with the $T$ axis will occur not at a large finite value $T_{γα} \sim 100$ K but at lower temperature, a long range magnetic ordering (for example AF) will occur at $T_N$. If a negative pressure could be realized (9), AF would appear at $T_N$ up to a pressure $P_C$ (see below the Doniach picture) where the ground state will switch from PM to AF. An obvious problem is what will be the consequence of the intercept between the valence transition line $(T, P)$ which must collapse at $P_{KL}$ at $T \to 0$ K and the AF-PM boundary at $P_C$ which may be slightly higher than $P_{KL}$ in Ce HFC. For simplicity, even for $n_f \sim 1$ an intriguing question is if the 4f electron must be considered as localized or delocalized i.e. takes part in the volume of the Fermi surface or not (image of small or large Fermi surface). Thus in addition to the magnetic instability at $P_C$, a possible Fermi surface instability may occur (for example at $P_{KL}$). Furthermore if $T_{CEP}$ gets comparable to any characteristic temperature $T_N$ for AF, $T_{Curie}$ for FM, $T_c$ for SC, the valence fluctuations at $P_v$ must be considered in addition to the spin instability at $P_c$.

The fascinating aspect is that very often the valence change is not marked by a clear first order line but corresponds to a smooth continuous crossover in $P$ and $T$. The physical reason is that in this complex matter, the f electrons are of course not the sole partner. The Coulomb repulsion $U$ between spin (↑) and spin down (↓) 4f electrons play a main role to stabilize a quasi-integer occupancy ($n_f \sim 1$). But also the Coulomb repulsion (term $U_{fc}$) must be considered in the interplay between the f and c electrons. That leads to the so called extended Anderson Hamiltonian with firstly the kinetic energy of the light electron (wavevector $k$, spin σ) and the term reflecting the position of the 4f level by respect to the Fermi energy $\varepsilon_f$:

$$H_0 = \sum_k \varepsilon_k c^+_{k\sigma} c_{k\sigma} + \sum_{i,\sigma} \varepsilon_f n^f_{i\sigma} + U \sum_{i=1}^N n^f_{i\uparrow} n^f_{i\downarrow}$$

and secondly the perturbation coming from the hybridization potential $V$:

$$V \sum_{i\sigma} \left( f^+_{i\sigma} c_{i\sigma} + c^+_{i\sigma} f_{i\sigma} \right)$$

where $c_{i\sigma}$ ($c^+_{i\sigma}$) is the annihilation (creation) operator of a conduction electron at the i site with spin σ and $f_{i\sigma}$ ($f^+_{i\sigma}$) is the corresponding term for an f electron and finally a $U_{fc}$ Coulomb repulsion term between the c and f electron on the site i:

$$U_{fc} \sum_{i=1}^N n^f_i n^c_i$$

The two first terms give the well known Anderson Hamiltonian; the third term is essential to derive V-FOT on sound basis. In the $U_{fc}$, $\varepsilon_f$ plane at $T = 0K$, the valence transition can be drawn with a first order line terminating now in a valence quantum critical point (V-QCP) (16-17) (figure 5) which is the extrapolation to $T = 0K$ of the critical end point. For most HFC, the V-QCP cannot be detected experimentally as it corresponds to a negative pressure or negative temperature value. However as it happen for all quantum singularities the electron can feel its position and then strongly reacts if under P or magnetic field (H) tunings it approaches the V-QCP location: that may lead to low energy fluctuations characterized by a temperature $T_v^*$.

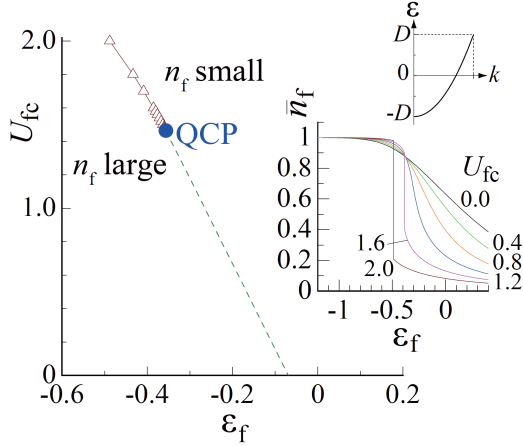

Fig. 5: Valence transition boundary at $T = 0K$, as a function of the parameter $U_{fc}$, $\varepsilon_f$ of the extended Anderson Hamiltonian. The first order transition at $T = 0K$, terminates at a QCP. At the left in the Kondo regime, $n_f \sim 1$, the valence is near 3; at the right $n_f < 1$, the valence deviates significantly from 3. The insert shows the light electron bandwidth and the variation of $n_f$ as a function of the position $\varepsilon_f$ of the 4f level for different values of $U_{fc}$ (17).

**4/ First contribution in the effective mass enhancement. The Kondo renormalization.**

A large part $m_k$ of the effective mass $m^*$ of the quasiparticle is gained through local fluctuations arising from the so-called exchange Kondo coupling $\Gamma \vec{S}.\vec{s}$ between the local spin S of the 4f electron and the spin s of the conduction electron. This many-body mechanism was initiated by the pioneer work of J. Kondo in 1964 (18) for a single paramagnetic impurity dissolved in a non magnetic host (e.g. Ce in $LaAl_3$) and solved completely by the Nobel laureate K. G. Wilson in 1974 (19). Its physical meaning had been clarified by P. Nozières with its local Fermi liquid picture (20) and K. Yosida and Y. Yamada (21). The novelty and the beauty of the effect is that a mechanism exists for a single paramagnetic impurity of spin S = ½ ( without interactions with the others ) to loose its magnetic entropy (Rln2) via its sole local strong coupling with the Fermi sea below a characteristic Kondo temperature $T_K$ (see figure 6):

$$T_K = \frac{1}{N(E_F)} \exp -\frac{1}{|\Gamma| N(E_F)}$$

where $N(E_F)$ is the density of states of the light electrons. The negative AF exchange constant $\Gamma$ is linked to the position $\varepsilon_f$ and the width $\Delta$ ($\sim V^2 N(E_F)$) of the 4f level according to the relation:

$$\Gamma N(E_F) = \frac{\Delta}{\varepsilon_f}$$

The characteristic Kondo temperature associated to the trivalent configuration can be expressed as a function of $n_f$, $\Delta$ and $N_f$ (the degeneracy of the 4f level) :

$$T_K = (1-n_f)\Delta N_f$$

In this hypothesis the susceptibility at 0K is $\chi \approx \frac{n_f}{T_K}$ (we use a slightly different definition of $T_K$ by linking its occurrence to the trivalent property). Going to the Kondo lattice, $T_K$ will be a first temperature scale which will govern the memory of the single impurity behaviour. Often, it is taken for the value of the effective Fermi temperature of the renormalized band which will be built by this many-body effect on the Fermi sea. Furthermore the strength of the Kondo energy plays a major role in the hierarchy and thus the efficiency of other processes such as the crystal field and the intersite magnetic interactions. For example if the Kondo energy $k_B T_K$ will be greater than the crystal field (CF) splitting $C_{CF}$ the full angular momentum $J$ of the 4f electron ($J = 5/2$ for $Ce^{3+}$, $J = 7/2$ for $Yb^{3+}$) will be already quenched by the Kondo fluctuations and the CF cannot lift further the degeneracy on cooling; $P_{CF}^*$ will mark the value of the crossover pressure where $C_{CF}=k_B T_K$. For Ce HFC, it happens always in the mixed valence phase; for Yb HFC, even for rather large departure of $n_f$ from unity ($v \approx$ 2.6), the Kondo temperature can be small enough to allow the lifting of the degeneracy in the mixed valence state. If $k_B T_K < C_{CF}$ the nice situation of a doublet crystal field ground state will be realized for the Kramer's ions like $Ce^{3+}$ and $Yb^{3+}$.

It is often considered that trivalent Yb ions with 13 electrons and thus one hole in the 4f shell are the 4f hole analogue of trivalent Ce with only one electron and 13 holes. However, there are two major differences (22):
i/ the deeper localisation of the 4f electron of Yb by respect to Ce leads that $\Delta_{Yb} \sim (1/10) \Delta_{Ce}$,
ii/ the larger strength of the spin orbit coupling ($\lambda$) between the $j = \ell - 1/2$ and $j = \ell + 1/2$ individual electron configuration for Yb than for Ce, $\ell = 3$ being the 4f orbital momentum. The consequence is that for Ce, the hierarchy is $\Delta > \lambda > C_{CF}$; as $k_B T_K$ cannot exceed the maxima of the energy involved in the problem, the equality $k_B T_K = \Delta$ requires that the occupation number $n_f$ stays above 0.84. For Yb, the hierarchy is reversed $\lambda > \Delta \sim C_{CF}$, $k_B T_K = \lambda$ can be achieved even with $n_f$ close to zero. Thus applying pressure on Yb compounds may allow scanning the valence from almost 2 to 3. For Ce HFC, the valence change under P will be restricted from 3 to 3.16. These hand waving arguments are supported by the recent band structure calculations shown in figure (7) for the $CeRh_2Si_2$ and $YbRh_2Si_2$ systems. It was also shown on multi-orbital model Anderson model that the existence of f-electrons on other orbital than the relevant one reduces considerably the coefficient B of the canonical formula of $T_K=B(1-n_f)\Delta$ (23).

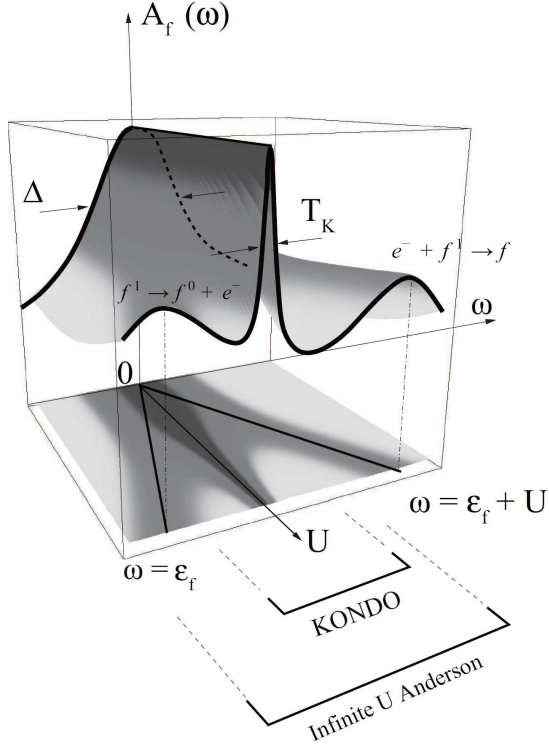

Fig. 6: Spectral weight $A_f(\omega)$ built by increasing the Coulomb potential $U$ between the spin up and spin down 4f electron with the continuous evolution from unresolved 4f levels strongly damped for $U = 0$ to broad levels plus a Kondo resonance (width $k_B T_K$) on switching $U$; the lower part is the density of the f spectral function where the non interacting resonance at $U=0$ slits into lower and upper level at $\varepsilon_f$ and $\varepsilon_{f+U}$ (37).

We have previously used local arguments to discuss whether the angular momentum is quenched or not before the CF will be operational (definition of $P_{CF}^*$). Similar statements have been made on band structure basis (24) by comparing the spin orbit splitting $\Delta_{SO}$ between the $j=5/2$ and $j=7/2$ bands, their bandwidth $\Delta_{BW}$ and their further splitting by switching $U$ in the case of no CF effect and of CF effect for quartet and doublet (see figure 8). In this last case, the Coulomb repulsion boosts the CF splitting in agreement with the fact that large $U$ helps to reach a low $T_K$ regime. The enhancement of $\Delta_{SO}/\Delta_{BW}$ is near 20 going from Ce to Yb. Thus novel phenomena must appear going from Ce to Yb HFC with notably remarkable quantum singularities directly linked with the possibility to select a low quantum state S=1/2 in the mixed valence phase. We strongly believe that this point is the main new window entering in Yb HFC territory.

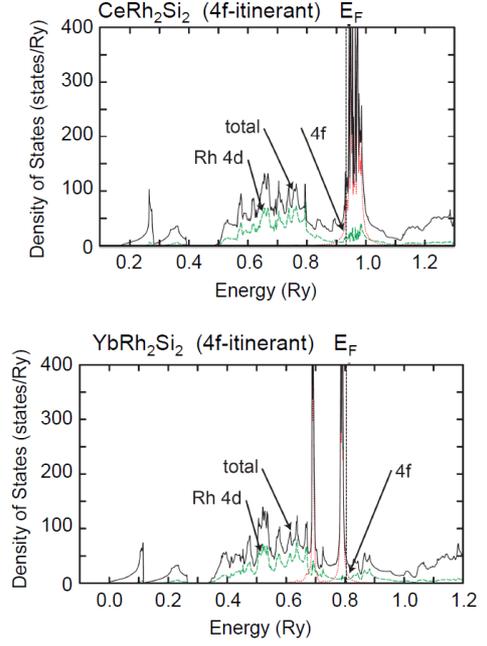

Fig. 7: Band structure calculations on $CeRh_2Si_2$ and $YbRh_2Si_2$ realized in the so-called local density approximation. We can see clearly for $YbRh_2Si_2$ the j = 5/2 and j = 7/2 subbands while for $CeRh_2Si_2$ they collides (22).

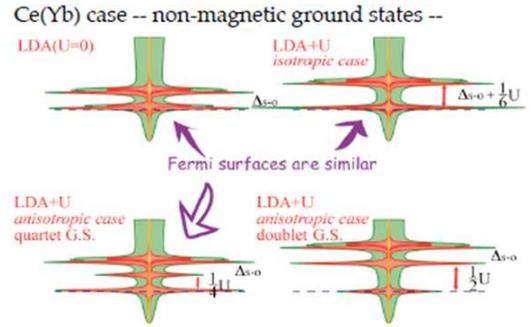

Fig. 8: Effect of U in LDA+U calculation taking into account an orbital dependent potential. When the bandwidth is small as it occurs for Yb HFC, U enhances CF and thus facilitates its efficiency.

Figure 9 presents an overview of the difference between Ce and Yb HFC. For Ce HFC, $P_{CF}^*$ is attached to $P_v$ where valence fluctuation or transition will be felt. As discussed later, $P_v$ is often a ghost pressure as the system will often only feel its proximity from valence transition. In a strong coupling case through the establishment of unconventional SC, $P_c$ well defined in the normal phase may also become a second ghost pressure. For Yb HFC, in agreement with the picture of electron hole symmetry between Ce and Yb cases, long range magnetism will be achieved at high pressure, $P^*_{CF}$ can be located deep inside the mixed valence phase ($n_f = 0.7$, $v=2.7$).

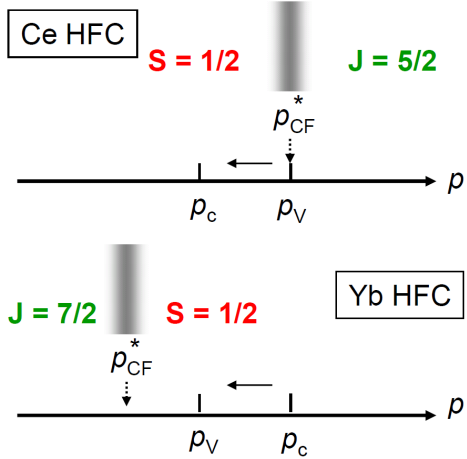

Fig. 9: Drastic difference between Ce and Yb HFC for the hierarchy between their characteristic pressure for the crystal field collapse ($P^*_{CF}$), for the valence transition ($P_v$) and for the magnetic instability ($P_c$).

## 5/ Thermodynamic measurements: evidence of other energy scale

In a classical picture of itinerant electrons, the Fermi temperature governs all electronic properties. For a single Kondo impurity, $T_K$ is the only relevant temperature. Macroscopic thermodynamic measurements are simple tools to test the validity of such an hypothesis by the simple comparison of the temperature dependence of the specific heat and of the thermal expansion $\alpha = \frac{1}{V}\frac{\partial V}{\partial T}$ which is of course directly related to the pressure derivative of the entropy (see the Maxwell relation: $\frac{\partial V}{\partial T} = -\frac{\partial S}{\partial P}$).

If the free energy $F$ can be written as a general unique function of $T/T^*$:

$$F = T\phi\left(\frac{T}{T^*}\right)$$

whatever in the temperature range, the thermal expansion and the specific heat will be proportional. That defines a unique Grüneisen parameter

$$\Omega_{T^*} = \frac{\alpha V}{C\kappa} = \frac{\partial \log T^*}{\partial \log V}$$

constant and independent of the temperature ($V$ and $\kappa$ are the molar volume and the compressibility). Reciprocally the ratio of $\alpha$ by $C$ will lead to define a Grüneisen parameter $\Omega(T)$; if it is strongly $T$ dependent the free energy cannot be written in the previous simple form.

For noble metal like Cu, the electronic contribution to $C$ and $\alpha$ gives $\Omega_{TF} \sim 0.66$ reflecting only the fact that the Fermi temperature $T_F$ varies as $V^{-2/3}$ and the effective mass $m^* \sim m_0$ is pressure invarian. For single Kondo impurity, $\Omega_{TK}$ is basically near 100 for $n_f \sim 1$ and around 20 for $n_f \sim 0.9$. For liquid He$^3$, considered often as a strong Fermi liquid, the mass renormalization from the bare $^3$He mass jumps from 3 to 6 as the pressure increases from zero to 34 bar on the melting curve. Even at this pressure, in the limit of $T \rightarrow 0$ K, $\Omega = -2$ remains low. The negative sign indicates that the enhancement of $m^*$ is dominated by the interactions between quasi-particles. Its relative small amplitude demonstrates clearly the non proximity to any magnetic instability for the liquid phase of $^3$He even on its melting curve (9).

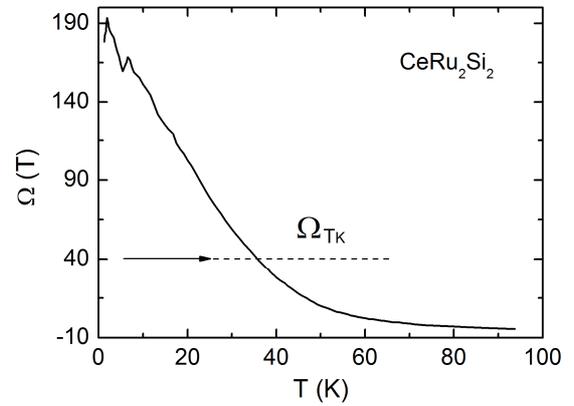

Fig. 10: T variation of the Grüneisen $\Omega(T)$ of CeRu$_2$Si$_2$ via the ratio of the 4f contributions of the thermal expansion by the specific heat. The large $T$ dependence of $\Omega^*(T)$ is a direct evidence of a non Fermi liquid regime (9,27). For a single Kondo impurity, $\Omega(T)$ = constant = $\Omega_{TK}$.

In heavy fermion compounds $\Omega(T)$ defined by the ratio of $\alpha/C$ is strongly temperature dependent and may reach a constant value only at very low temperature. Even for global criticality (see below), $\Omega(P_c)$ is predicted to diverge at $T \rightarrow 0K$ (25 ) In CeRu$_2$Si$_2$ (located few tenth of GPa above $P_C$), $\Omega^*(0) \sim 200$ (26) and as see in figure 10 a constant value is achieved far below the temperature $T_K \sim 20$ K attributed to its single impurity Kondo behaviour. Thus another characteristic temperature $T^*$ linked to magnetism or valence fluctuations governs the low temperature behaviour of heavy fermion compounds. Above $T^*$, the combined effect of strong quantum and thermal fluctuations lead to a so-called non Fermi liquid regime. In this temperature domain for example the specific heat does not follow the usual behaviour with linear $\gamma T$ term with additional $T^3$ or $T^2 \log T$ contribution but is dominated by a huge $\log T$ increase on cooling or the resistivity does not obey a $AT^2$ dependence but varies often with a power law $T^n$ with $n$ often close to 1. One of the debates is the origin of this non Fermi liquid behaviour. Does it require a complete revision of the Landau approach of interacting electrons?

Often in HFC close to $P_C$, $\Omega(0)$ and $m^*$ have qualitatively rather similar pressure dependence (27): an increase of $m^*$ under pressure is associated to an increase of $\Omega(0)$. If the ratio $\Omega(0)/m^*$ will be constant, the interaction will vary as $\log r$ ($r$ being the distance between Ce or Yb sites). Thus it implies long range couplings. In fact $\Omega(0)$ varies often more rapidly under pressure than $m^*$. The divergence of $\frac{dT}{dP}$ at the quantum phase transition at $T \to 0K$ is well known for first order phase transition (FOT). The Clapeyron relation $\frac{dT}{dP} = \frac{\Delta V}{\Delta S}$ which links the entropy ($\Delta S$) and volume ($\Delta V$) drop at the critical temperature requires that $\frac{dT}{dP} \to$ infinity since due the third thermodynamic principle $\Delta S$ will be zero. Only a volume discontinuity can occur at $T=0$ via a first order transition (figure 11). Thus it is not a trivial task to differentiate between a second order and a first order transition. Of course if volume or sublattice magnetization discontinuities are detected as well as evidences of a phase separation, the first order nature of the quantum phase transition will be clarified. Recent thermal expansion experiments under pressure show that in HFC even for a well defined first order transition the relative volume discontinuity is often near $10^{-4}$ i.e. three order of magnitude lower than that detected at $T_v$ in Ce metal or for $He^3$ on its melting curve (9).

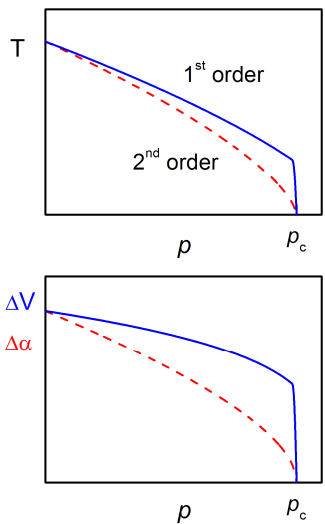

Fig. 11: Typical variation under pressure of the ordering temperature for a first (solid lines) and second (dashed) order transition with associated effects on volume discontinuity $\Delta V$ or thermal expansion ($\Delta \alpha$) anomaly.

An interesting case appears when the transition changes under pressure from second order to first order near the quantum singularity as it seems now to be established for the ferromagnetic quantum instability, with the occurrence of a tri-critical point (TCP) (28) detected via the convergence of three lines in a additional scan in magnetic field (figure 12). For an AF quantum singularity at $P_C$, the debate on the second order (quantum critical point) or first order nature of the singularity is still open.

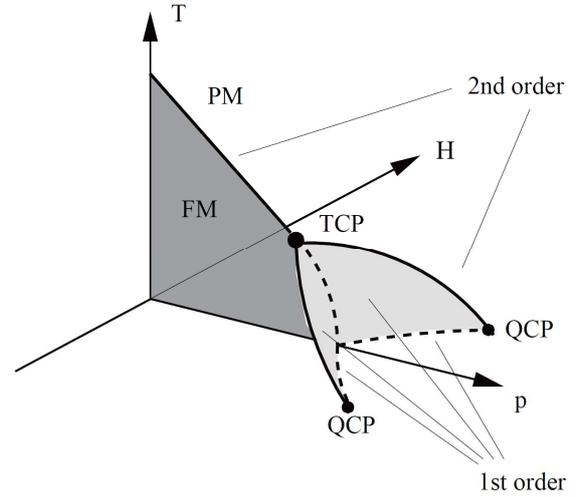

Fig. 12: Typical ($T$, $P$, $H$) predicted for FM itinerant systems with a TCP and two QCP (28).

**6/ Close to magnetic instability: global or local criticality**

In the classical case of long range magnetic ordering at a finite temperature $T$ (at $T_N$ for AF), a large specific heat anomaly marks the phase transition; the magnetic coherence length $\xi_{coh}(T)$ diverges at $T_N$ (figure 13) and decreases again on cooling down to an atomic distance $a$ in agreement with the fact that the interactions are limited to first neighbours. Applying pressure may change the magnetic structure but, at $T=0$ K, $\xi_{coh}(0) \sim a$ remains restricted to a first neighbour distance. At a quantum phase transition where $T_N$ collapses to zero, $\xi_{coh}(0)$ will diverge at $P_C$; the recovery of a coupling strength limited to an atomic distance will be achieved only far above or far below $P_C$. What are the main trends for the description of the magnetism collapse?

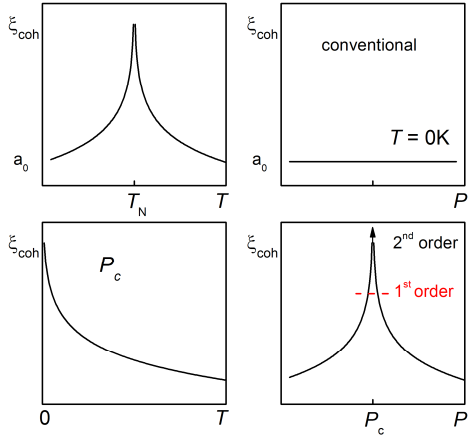

Fig. 13: Schematic variation of the magnetic coherence length as a function of $T$ and $P$. If the phase transition is first order, there will a cut off of the magnetic coherence length.

**The rigid Doniach scheme**

The first approach of the treatment of the magnetic instability of HFC was made by S. Doniach for Ce Kondo lattice (29). The first idea is to start with the image of a single Kondo impurity (energy $k_B T_K$ favourable to end up in a non-magnetic ground state) and the second step is to add an intersite magnetic interactions $E_{ij} \sim \Gamma^2 N(E_F)$ transmitted by the Ruderman Kittel Kasuya Yosida oscillations. As $T_K$ has an exponential dependence, the Kondo energy will rapidly overpass the RKKY interaction under pressure and thus long range magnetism will disappear. Using a simple relation for $\gamma$ derived from the field effect on the Kondo phenomena:

$$\gamma = \frac{k_B T_K}{(k_B T_K)^2 + (g\mu_B H)^2}$$

where $g\mu_B H$ is the Zeeman energy of a spin ½ and replacing $H$ by the molecular field $H_m$; it is easy to see that the effective mass $m^* \sim \gamma$ of the quasiparticle will reach a maximum at $P_C$ when $H_m$ collapses. For $P \ll P_C$, the Kondo mass enhancement $m_K \sim 1/T_K$ will be strongly destroyed by the large strength of $H_m$ and for $P \gg P_C$, the $P$ decrease of $m^*$ is governed by the pressure increase of $T_K$ (figure 14).

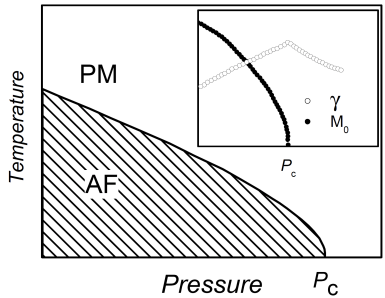

Fig. 14: Typical Doniach phase diagram for a Kondo AF. $T_N$ ($P$) will collapse at $P_C$; the insert corresponding pressure variation of $\gamma$ and of the sublattice magnetization $M_o$ (9).

For Yb at first glance the situation will be symmetrical. However in Yb HFC, the interplay between valence and spin instability is suspected to be strong as for the same strength of $T_K$, the departure of $n_f$ from unity will be an order of magnitude greater in the Yb case than in the Ce case. Another view (30) on their differences in a P response is that, if for both elements $\Delta$ must increase under pressure as governed by the increase of the light electron bandwidth, for Ce $|\varepsilon_f - E_F|$ will decrease under $P$ while for Yb $|\varepsilon_f - E_F|$ will increase under $P$ leading to a far moderated $P$ variation of $\Delta/|\varepsilon_f - E_F|$ then of $T_K$ in Yb HFC than in Ce HFC. A nice illustration is given by the comparison of the pressure variation of $T_N$ observed for CeRh$_2$Si$_2$ and YbRh$_2$Si$_2$ (22) (figure 15).

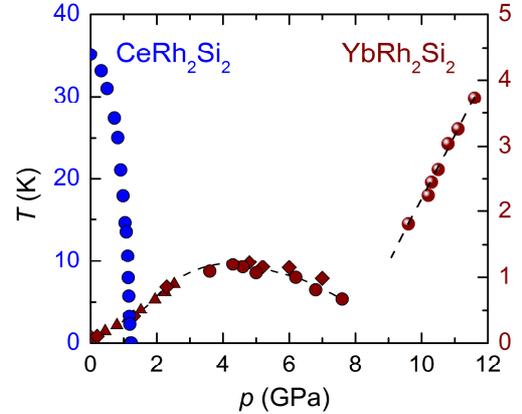

Fig. 15: Pressure dependence of $T_N$ of CeRh$_2$Si$_2$: the collapse of AF occurs in a narrow pressure window. For comparison, the $P$ dependence of $T_N$ in YbRh$_2$Si$_2$ is shown. The slow $P$ evolution of $T_N$ was taken here as an evidence of the interplay between magnetic and valence fluctuations (22).

**Global criticality: the spin fluctuation view**

In the previous picture, the magnetism is treated as a rigid quantity; the many body effect is introduced via the Kondo concept which may become irrelevant for the lattice at very low temperature. As discussed near $P_C$, large coherence length must occur i.e. basically an extra-correlation length $l_{KL}$ must appear. Assuming that the quasiparticle of effective mass $m^* \sim 1/T_K$ circulated along a Kondo loop of length $l_{KL} \sim (m^*/m_0)a$, their lifetime $\tau_{KL} \sim m^{*2} \sim T_K^{-2}$ will open a new energy scale $k_B T_{KL} \sim (\tau_{KL})^{-1} \sim T_K^2$. Thus a new energy scale must occur when the coherence length overpasses strongly the interatomic distance $a$ (9).

To take into account the key role of longitudinal and transversal spin fluctuations near the AF wavector $Q_0$, a sound approach was to extend the powerful spin fluctuation theory derived for itinerant 3d electrons in metals by assuming that the quasi-particles are already renormalized locally by the Kondo effect ($m_K$) which will play the role of the renormalized band mass ($m_B$) and further enhanced via AF or FM spin fluctuations ($m^*/m_K$) (see 25). This theory predicts that on approaching $P_C$, the so-called Fermi liquid domain will shrink below a spin fluctuation temperature $T_A \sim (P-P_C)$ (figure 16). As pointed out near V-QCP, now near M-QCP two characteristic temperatures $T_K$ and $T_A$ will govern the temperature behaviour of the physical properties. Between $T_K$ and $T_A$ the large non Fermi liquid regime characteristic of a crossover regime will occur. The interesting point is that close to $P_C$ the electron will feel the proximity of the quantum singularity even at high temperature (far above $T_A$) as if it knows its destiny. The same remark is true for other quantum singularities as CEP or TCP. Furthermore the nature of the singularity will play a main role in the $T$ crossover law. Experimentally $T_A$ will be defined by the temperature below which Fermi liquid properties will be recovered such as a $AT^2$ law for the resistivity; as we will see it is not obvious if $T_A$ must be associated to spin or valence fluctuations without direct microscopic evidences. In the AF spin fluctuation approach (25), $\gamma \sim m^*$ has a singularity in $\sqrt{P-P_C}$ at $P_C$ which corresponds to a sharp maximum but not a divergence of the effective mass; however $\Omega_{Pc}$ at $T=0K$ must diverge as $(P-P_c)^{-1/2}$.

For FM interactions $\gamma \sim -\log(P-P_C)$ and thus a divergence of $m^*$ will be expected at $P_C$ so far a real QCP exists i.e the FM sublattice magnetization collapses continuously at $P_C$. Such a divergence does not occur as the system prefers to end up its FM via a first order transition. There is now a theoretical and experimental consensus that the FM quantum singularity is first order. Two underlying hypotheses of the spin fluctuation approach are the invariance of the FS through $P_C$, the focus on a well defined wavevector $Q_0$ for the magnetic instability which will induce a global instability.

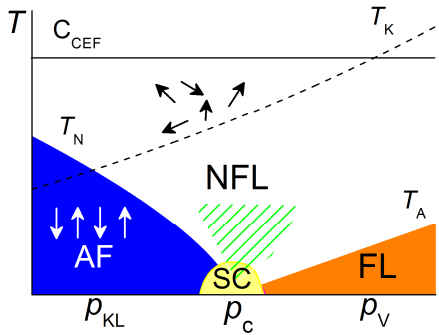

Fig. 16: The spin-fluctuation scheme (global criticality) assumes the invariance of the Fermi surface through $P_C$ and gives a simple explanation on an extended non Fermi liquid (NFL) regime for $P \sim P_C$ and on the collapse of the characteristic spin fluctuation temperature $T_A$ at $P_C$.

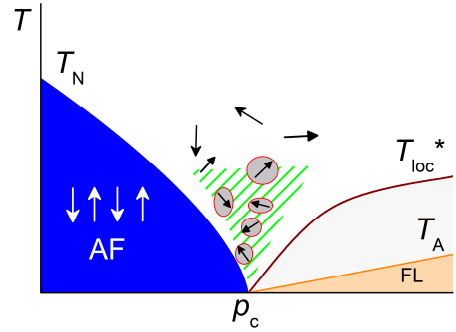

Fig. 17 The local criticality stresses the importance of Fermi surface reconstruction with the image of small Fermi surface in the AF phase and large Fermi surface in the PM regime. The modification of the FS may be achieved through a crossover regime at $T_{loc}^*$. The previous Kondo line of fig. 16 has been wiped out with the idea of a Kondo breakdown near $P_C$.

**Local criticality**

Inelastic neutron scattering experiments (31) support the spin fluctuation image for some cases as the CeRu$_2$Si$_2$ series but seem to rule out this approach as a universal rule. For example in the CeCu$_{6-x}$Au$_x$ series, the magnetic instability near $P_C$ appears to involve all the wavevector response (32). That leads to derive models of so-called local criticality (instead of the previous global criticality) which stress that the Fermi surface (FS) is instable at $P_C$ (image of small and large Fermi surface) (33-34) (figure 17) with the appearance of a new characteristic temperature referred as $T_{loc}^*$. In the previous global criticality scheme, $T_{loc}^*$ will collapse for $P$ lower than $P_c$ (reminiscent of our $P_{KL}$ point). In the local criticality scenario, $T_{loc}^*$ is assumed to collapse just at $P_c$. In this case, the quantum critical fluctuations are not restricted to the fluctuation of the magnetic order parameter but also to extra soft variables generated either by fermionic hole excitations or other degrees of freedom like those associated with the Kondo effect (image of Kondo breakdown or Mott localisation). As pointed out (12), a full understanding of coupled soft modes is not available despite a variety of proposals. We underline later that valence fluctuations must be considered more seriously. Experimentally there are cases where directly via dHvA experiments there are no doubt that the FS changes at $P_C$ like in CeRh$_2$Si$_2$ and CeRhIn$_5$ where excellent agreements are found between dHvA frequencies detected below $P_C$ with band structure calculations assuming the 4f electrons localized and detected above $P_C$ with band structure calculations assuming the 4f electrons itinerant. An experimental difficulty in an unambiguous Fermi surface determination via the dHvA quantum oscillation technique is that it requires large magnetic fields to see the motion of the electron on their quantized orbits while the quantum phase transition itself is often quite sensitive to the magnetic field. Unfortunately high energy spectroscopy experiments like photo-emission or ARPES (angular resolved photoemission spectroscopy) are very often limited to rather high temperature (few Kelvin) by comparison to quantum singularity problems; however, great progresses have been realized in this direction. Detailed discussions on the duality between global and local criticality can be found in references (12,35,36,37).

**7/ The magnetic field scan: quantum criticality under magnetic field**

Tuning by the magnetic field gives a further opportunity to feel quantum singularities. For the magnetic singularity as M-QCP at $P_C$, depending on the spin and interaction anisotropy quite different conditions will occur (figure 18). For Ising spin type systems like $CeRu_2Si_2$ or $CeRh_2Si_2$, the observed metamagnetic transition at $H_C$ from AF to PM ends up at a finite CEP at $H_c^*$ (figure 18a); entering in the PM domain above $P_C$ this CEP is felt even far above $P_C$ as shown by the occurrence of sharp pseudo-metamagnetic phenomena at $H_M$ which shows up sharp magnetostriction effects and strong maximum in the enhancement of the effective mass. The particularity is that this metamagnetic phenomenon occurs for a constant value of the magnetization. $H_M$ cannot be distinguished often from the crossover field $H_K = \dfrac{k_B T_K}{g\mu_B}$ above which a slow recovery of the full magnetization of the $Ce^{3+}$ configuration will happen.

When the spin anisotropy is weak (figure 18b), $H_C$ is not associated with metamagnetic phenomena but is connected to the continuous closing of the $H$ misalignment of the sublattice magnetization. As $H_C$ will collapse in the vicinity of $P_C$, $H_C$ and $H_K$ are clearly decoupled. Such a phenomenon is observed in the $CeNi_2Ge_2$ series (see 22) and as we will see later in $YbRh_2Si_2$ (see 9).

Of course when the magnetic polarization becomes high, the spin up Fermi Surface will expand at the shortage of the spin down one. If it touches the Brillouin zone, a Fermi surface reconstruction (38-39) will occur with the fancy problem that the minority spin down electron may be so heavy that their Fermi surface is completely hidden; the heaviness of the minority spin coming from their difficulty to move in such dense majority spin medium (local repulsion $U_{ni\downarrow ni\uparrow}$). The majority spin up Fermi surface will look rather similar to that calculated for localized 4f electrons as their heaviness is strongly depressed. The debate between the localization above $H_M$ of the 4f electrons or the persistence of its itinerancy is still open in the absence of a complete determination of the Fermi surface and also of reliable band structure calculations in magnetic field which are extremely difficult. A highly documented case is that of $CeRu_2Si_2$ where at least the Fermi surface is totally determined below $H_K$ (40,41).

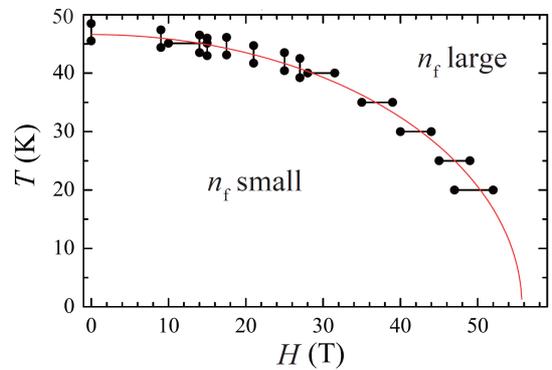

Fig. 19: Doping with La and Th in a $Ce_xLa_{1-x}Th_{1-x}$ compound allows to lower $T_{\gamma\alpha}$ ($P = 0$)= from 120K to 48 K. Value sufficiently low to be decreased continuously to zero for a field of $H = 56$ T (42).

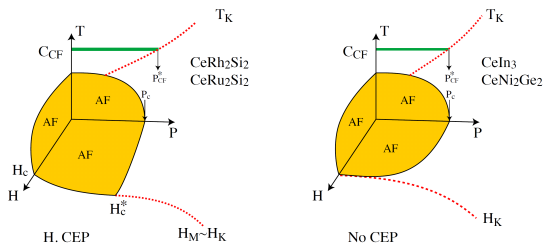

Fig. 18: (left frame) ($H$, $P$, $T$) phase diagram of an Ising type Kondo lattice. The magnetic critical field $H_C$ of the AF → PM metamagnetic transition ends up at a finite value $H_c^*$, an important point will be the pressure $P \sim P_{CF}^*$ where the crystal field effect will be wiped out by the strength of the Kondo energy $k_B T_K$. (right frame) ($H$, $P$, $T$) for 3 dimensional Heisenberg type Kondo lattice $H_C$ will collapse with $T_N$ at $P_C$. However reaching high magnetic polarisation near $H = H_K$ may lead to feel V-QCP or to FS reconstruction between majority and minority spin.

A new discussed possibility is the H action on the valence transition $T_v$. In the case of the first order transition $T_v(P)$, the magnetic field can have a strong effect as it can drive $T_v$ near the other characteristic temperature like $T_N$ or $T_C$ and thus leads to a strong interplay between the phenomena. According to the Maxwell relation $\left(\dfrac{\partial V}{\partial H} = -\dfrac{\partial M}{\partial P}\right)$ for both cases of Ce and Yb elements, the magnetic field favours the restoration of the $3^+$ configuration: for Ce an expansion of the volume and for Yb a contraction of the volume. Evidences of a field collapse of $T_v(0)$ was first pointed out experimentally on the Ce$_{1-x}$La$_x$Th$_x$ (42), where the substitution decreases $T_{\gamma\alpha}$ at $P = 0$ (figure 19). In agreement with a crude model (43) based on isolated local entropy (consideration on $T_K$), the field variation of $T_v$ satisfies the relation:

$$\left(\frac{H}{H_v}\right)^2 + \left(\frac{T}{T_v}\right)^2 = 1$$

where $H_v$ is the field value for the first order valence transition.

The same phenomena were beautifully demonstrated for the Yb system YbInCu$_4$ (44). In figure (20) which represents $T_v(P)$ of YbInCu$_4$ one can recognize the electron-hole symmetry between YbInCu$_4$ and the previous Ce phase diagram. When $T_v(P)$ becomes low enough near 2.5 K, first order long range magnetic ordering (45-46) is detected with a phase separation window around 2 GPa between the PM and the FM phase. It is interesting to point out that it was recently suggested that near a collapsing valence transition, FM will be favoured via the occurrence of large non local dynamical deformation of the lattice (47). Microscopically the field dependence of $T_v$ ($H$) at $P=0$ was even observed via X-ray absorption spectroscopy. Figure 21 shows the field variation of the valence which is near 2.84 at $H = 0$ and reaches 2.96 for $H = H_v$ (48).

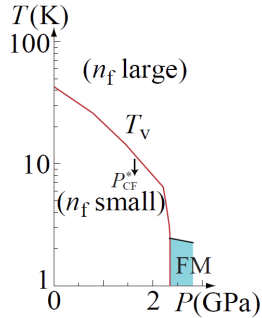

Fig. 20: Schematic ($T, P$) phase diagram of YbInCu$_4$. $T_V$ has been determined either by resistivity and NQR, $T_{Curie}$ by susceptibility and NQR. The arrow indicates the location of $P_{CF}^*$ (45, 46, 65)

Recently the theoretical study on the field effect on the V-FOT (16-17) was extended to the mechanism in which way the critical end points of V-FOT are controlled by a magnetic field. The results at H = 0 presented previously were now also calculated in magnetic field. As seen in figure (22), the CEP can change drastically under magnetic field and furthermore its $P$ dependence leading to change of $\varepsilon_f$ and $U_{fc}$ can lead to a marked minimum. As pointed out, few V-FOT are observed; however in many cases, the 4f electron will feel its proximity to CEP. Figure (22) suggests that in the crossover valence regime close to CEP, the corresponding crossover temperature $T_V$ can go to a maximum in field by contrast to the continuous decrease of $T_v$ with $H$ for the V-FOT (see later YbRh$_2$Si$_2$).

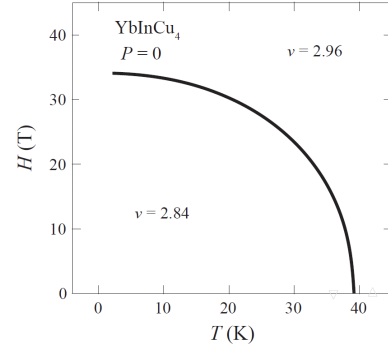

Fig. 21: At $P = 0$, $H$ dependence of $T_V$ ($P = 0$) as determined by X ray spectroscopy (48).

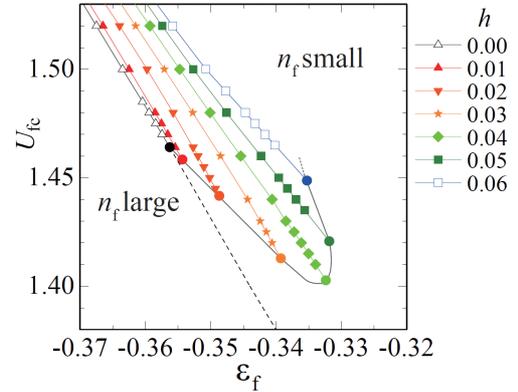

Fig. 22: Calculation of the magnetic field effect on the valence transition in the extended Anderson Hamiltonian (16-17).

Thus for the magnetic properties, it may happen that $T_v$ and $H_v$ or its corresponding crossover value becomes comparable to $T_N$ and $H_C$. Thus two bare phase diagrams will overlap with interesting interplays between their quantum singularities. It is worthwhile to notice that in figure 18, considerations between a phase transition of the Kondo lattice and the single impurity behaviour ($T_K$, $H_K$) have been mixed. A sound approach is to compare $H_V$ and $T_V$ with $H_C$ and $T_N$. $T_K$ and $H_K$ are single impurity parameters which will control the efficiency of local perturbations like the crystal field splitting or the magnetic polarization. Of course the pressure dependence of $T_K$ and $H_K$ is the driving factors to define the magnetic and valence ($T, H, P$) phase diagram. To our knowledge, there is yet no complete theoretical works treating valence transition with the possibility of change in the crystal field degeneracy; for Ce HFC, theoretical arguments can be found in ref 49 (pages 8-9). The interplay between two phase diagrams belongs to the exotic classes of $H$ re-entrant phenomena where $H$ re-entrance of one phase occurs in the other high pressure stable phases such as recently discovered for AF in the SC phases of CeRhIn$_5$ (50-51).

## 8 / Current research areas in HFC

Among recent research areas in HFC, one can list:

- the continuous attempt to solve the Kondo lattice problem with its different quantum singularities (32-33),
- many observations of non Fermi liquid behaviour close to $P_C$ (12-35),
- new attempts to clarify other quantum singularities like the QCP : CEP, TCP,
- the precise determination of the ($T$, $P$) phase diagram with special attention on the interplay between magnetic or valence instabilities and unconventional SC (49,52, 53),
- magnetic field studies of magnetic phase transition including scans through $P_C$, $H_C$ (9,12, 35),
- magnetic field studies at the interplay of SC and AF or FM (50,51),
- careful determination of Fermi surfaces on both side of $P_C$, $P_V$, $H_C$, $H_v$ (54),
- rush to new cases such as discussed above with the discovery of new Yb materials and recently with SC in Ce compounds with non-centrosymmetric lattices (55-57)

In the field of highly correlated electronic systems, the real boost in the last three decades was coming from the discovery of unexpected properties of new materials: high $T_{SC}$ oxides (6), two band SC in MgB$_2$ (58) and recent pnictide superconductors (59). The optimistic view that new structural arrangements must lead to new effects has been often rewarded. In the field of HFC, the growth of excellent crystals of the so-called 115-family (CeRhIn$_5$, CeCoIn$_5$, CeIrIn$_5$) where Ce planes of CeIn$_3$ are intercalated by a $T$In$_2$ plane ($T$: transition metal) (60), allows precise experiments of the interplay between AF and SC in pressure and magnetic field over a large temperature window as the maxima in the respective critical temperatures (maximum of $T_N \sim 3.5$ K, maximum of $T_c \sim 2.4$ K) are firstly quite similar and secondly quite favourable for a deep incursion in the low temperature regime (large $\frac{T_N}{T}$ or $\frac{T_C}{T}$ exploration).

Furthermore, the corresponding critical fields ($H_C <$ 40 T and $H_{C2} <$ 12 T) are quite accessible experimentally ($H_{C2}$ is the upper critical superconducting field above which the normal phase is recovered). HFC are key systems to study the link between unconventional superconductivity and quantum singularities due to the opportunity to work on clean crystals which are easy to grow by competing teams. This stimulating atmosphere has allowed a major breakthrough in the occurrence of unconventional SC. Unconventional SC occurs through the interactions between quasi-particles mediated via spin or valence fluctuations or via crystal field excitations. Before the discovery of the 115 - family, the main link between unconventional SC and quantum singularities was to recognize that maxima of $T_C$ appears close to a quantum singularities (M-QCP or V-QCP) (see CePd$_2$Si$_2$ on figure 23). For CeRhIn$_5$ the ($P$, $T$) phase diagram has been carefully determined by thermodynamic measurements such as microcalorimetry (50-51) or by NQR or NMR spectroscopy (61) and additionally by neutron scattering experiment (62,63). Figure 23 illustrates the results of such studies in the ($P$, $T$) phase diagram of CeRhIn$_5$.

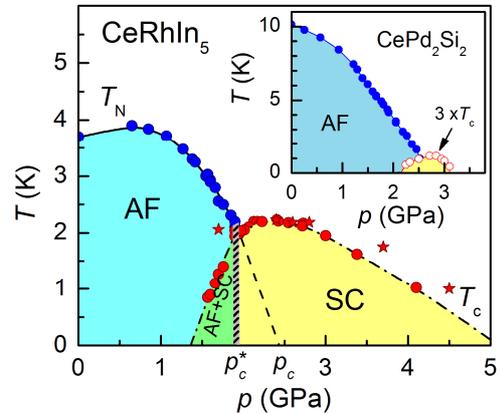

Fig. 23: The comparable values of the maxima of $T_N$ and of $T_{SC}$ open the possibility to discuss the interplay between magnetism and SC with many facets: repulsion, coexistence, field re-entrance. This possibility was quite impossible in the case of CePd$_2$Si$_2$ where $T_N$ (max) $\sim 10$ K and $T_{SC}$(max) $\sim 0.4$ K (9, 51).

In comparison we show the phase diagram of CePd$_2$Si$_2$, the low value of $T_c$ makes it quite difficult to perform detailed microcalorimetric experiments. We will not enter here in the physical discussion of the ($P$, $T$) phase diagram of CeRhIn$_5$ and notably on the domain where AF and SC coexists homogeneously at low temperature but stress again that the proximity of $H_C$ and $H_{C2}$ leads to the remarkable phenomena of the field re-entrant AF which is glued to SC.

Up to now most studies on HFC are concentrated on the Ce or U examples, the physical reason being that the rather large value of $\Delta$ for Ce facilitates the "fishing" of quantum singularities. The research target on Yb HFC is to point out new phenomena linked to the sharpness of their internal structure ($\lambda > \Delta \sim C_{CF}$) with clear signatures of the separation of the interplay between competing effects. One difficulty to grow high quality Yb compounds is its high vapour pressure, but improvements of the flux method open now the possibility to grow excellent Yb crystals. We restrict our description to four cases: (i) YbInCu$_4$ and its associated YbXCu$_4$ with the previous described V-FOT and with special attention on the YbAuCu$_4$ case which may be rather similar to the highly discussed example of YbRh$_2$Si$_2$, (ii) YbCu$_2$Si$_2$ shows a valence crossover which ends up at high pressure in a magnetic phase, (iii) the exotic YbRh$_2$Si$_2$ example with its nice non-Fermi liquid singularities, and (iv) β-YbAlB$_4$ where for the first time Yb HFC-SC has been discovered.

**9/ New windows the Yb heavy fermion compounds: hope and present status. Four Yb HFC series: YbInCu$_4$, YbCu$_2$Si$_2$, YbRh$_2$Si$_2$ and β-YbAlB$_4$**

The valence of these four Yb HFC has been evaluated by high energy spectroscopy. For the four systems, it varies between $v = 2.7$ and $v = 2.9$; a precise absolute determination of $v$ is difficult but a reference can be made at least to the high pressure range where a trivalent state will be approached. Let us look to YbRh$_2$Si$_2$ which appears at $P=0$ to be closer to an Yb$^{+3}$ configuration than the other three examples. Recently (64) it has been shown by resonant inelastic x-ray scattering experiments at 300 K that a fully localized Yb$^{3+}$ state is only realized at high pressures of about 8.5 GPa and that at $P=0$ the valence is around 2.9. Of course following the Anderson impurity model and also the proximity to a valence singularity the valence at $T = 0$ will be even lower. So for the case of YbRh$_2$Si$_2$, the non-integer value of the valence must play an important role even if the V-FOT is hidden.

**YbInCu$_4$ : from FOVT to proximity of V-CEP (YbXCu4 with X = Ag, Cd ) and then appearance of AF (YbAuCu$_4$)**

By comparison to the three latter cases of YbCu$_2$Si$_2$, YbRh$_2$Si$_2$ and β-YbAlB$_4$ the great interest in YbInCu$_4$ is that V-FOT is observed with $T_v$=45K at $H$=0. An interesting feature is that CF effects ($C_{CF}$ around 30 K) are detected only above $P_{CF}^* = 1.2$ GPa (65). The reason of the CF collapse is that at low pressure the Kondo temperature of the trivalent configuration for its full 2J+1 degeneracy is too high near 400K. We have already discussed the textbook example of V-FOT of YbInCu$_4$ with even the collapse of $T_v$ ($H$) under field at $P = 0$ for $H = 40$ T (fig 20- 21). Near 2 GPa, long range magnetic ordering (45-46) occurs in a regime of a strong interplay between valence and spin fluctuations. As at the borderline $T_v$ is finite the appearance of long range ordering (mainly FM) corresponds to a M-FOT. This statement is reinforced by a pressure window of coexistence between PM and FM phases. When In is replaced by Ag or Cd (YbAgCu$_4$ and YbCdCu$_4$ systems), no V-FOT appears as well as no magnetic transition can be detected. While the Kondo temperatures of both materials ($T_K$ ~ 200 K) are similar, only YbAgCu$_4$ shows a metamagnetic behaviour in the magnetization curve confirming that the vicinity form CEP depends of tiny differences in the bare parameter ($\varepsilon_f$, $U_{fc}$, $V_{cf}$) (see 16,17).

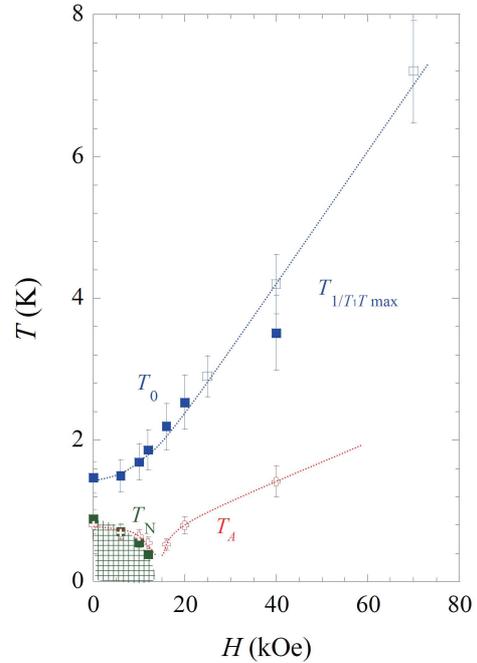

Fig. 24: ($T$, $H$) phase diagram of YbAuCu$_4$ at $P = 0$. $T_0$ corresponds to the temperature maxima of $1/T_1T$ ($T_1$ is the nuclear relaxation time) $T_A$ is derived from resistivity (66).

The intriguing case is that of YbAuCu$_4$ (66) as it orders below $T = 0.8$ K with a residual γ term (γ ~ 150 mJmole$^{-1}$K$^{-2}$) and a critical field $H_C$ ~ 1.3 T. As we will see later it reproduces a situation rather similar to that described below for YbRh$_2$Si$_2$ where $T_N$ ~ 0.07 K, γ ~ 1200 mJ mole$^{-1}$K$^{-2}$ and the critical field $H_C$ ~ 0.06 T for the easy magnetization plane i.e basically the same property than those of YbAuCu$_4$ but with all characteristic parameters droped ($T_N$, $H_C$) by a factor 10. Like in YbRh$_2$Si$_2$, the temperature $T_A$ below which Fermi liquid $AT^2$ law is obeyed in the resistivity at constant field cuts the $T_N$ ($H$) phase diagram at a finite value (fig. 24). Knowing the previous results on the YbInCu$_4$ series, it is appealing to consider that the properties of YbAuCu$_4$ result from the interplay between magnetism and valence transitions.

**YbCu$_2$Si$_2$: at *P* = 0, an old case of intermediate valence character ($n_f$ ~ 0.8) and strong crystal field effect**

Looking two decades ago, YbCu$_2$Si$_2$ provides certainly the best illustration that, in Yb HFC CF effects can be efficient to lift the 4f degeneracy originated in the trivalent configuration; its valence was estimated near $v$ = 2.8 ($n_f$= 0.8) (67,68); thus at *P*=0 it is considered to be in the small $n_f$ domain, i.e. in an intermediate valence state. At zero pressure, the CF effect was clearly observed: (i) strong anisotropy of the susceptibility with $\chi_\parallel$ to the c axis near three time $\chi_\perp$ at *T* = 0 K and of the Knight shift measured on the Cu nuclei (69); (ii) unusual *T* dependence of the 4f quadrupole moment (70) which can be well explained in a theoretical model (70) with a Kondo temperature in qualitative agreement with the estimation determined by specific heat data ($T_K$ = 50 K) and with an CF scheme derived from in-elastic neutron scattering experiments (71) (four doublet CF levels at 0, 18, 23 and 31 meV with a ground state formed mainly by the ±7/2 state) (72). From the temperature *T* = 90K where thermal dilatation (70) reaches its minimum a crossover valence temperature around 90K can be estimated. Above 200K, $\chi_\parallel$ and $\chi_\perp$ follow a Curie-Weiss behaviour with respective Weiss temperatures of -26 K and -157 K; on cooling at low temperature the striking point is that $\chi_\parallel$ increases while $\chi_\perp$ goes through a maximum at *T* ≈ 40 K. This difference implies via the Maxwell relation, $(\partial S/\partial H)_T=(\partial M/\partial T)_H$, and its *T*-derivative, $(\partial(C/T)/\partial H)_T=(\partial^2 M/\partial T^2)_H$ that applying a magnetic field parallel or perpendicular to the c axis will induce either an initial decrease or increase of the effective masses. It suggests a large anisotropy in the quasiparticle interactions with competing ferromagnetic and antiferromagnetic channels. Such phenomena are of course absent in the previous description of H-VQCP as the CF effects are omitted. We use this opportunity to stress that the territory of *H, P* scan will be very rich. The case of YbRh$_2$Si$_2$ described below will be an excellent example.

Improvements in flux growth methods (73,74) allow producing high quality crystal of YbCu$_2$Si$_2$. Thus it opens the possibility of a direct determination ot the Fermi surface as well as a new focus on the macroscopic low temperature behaviour. Combined studies of dHvA oscillations and band structure calculations (73-75) have been realized for the two cases of YbCu$_2$Ge$_2$ and YbCu$_2$Si$_2$ where *v* is respectively 2 and 2.8. For YbCu$_2$Ge$_2$, the 4f bands are located deep below the Fermi level and do not contribute to the Fermi surfaces while for YbCu$_2$Si$_2$, the 4f electrons appear as itinerant. High effective masses up to 30-40 $m_0$ have been detected in good agreement with the rather large electronic specific heat coefficient $\gamma$ = 150 mJ mole$^{-1}$K$^{-2}$. Despite the initial value of $n_f$ ~ 0.8 at *P*=0, microscopic probes such as Mössbauer spectroscopy (76), and macroscopic probes (as resistivity (77), microcalorimetry and susceptibility measurements (74)) lead to the conclusion that long range magnetic (mainly FM) ordering will appear only near 7.5 GPa. As indicated on figure 25, a finite pressure range (what is a mark of possible phase separation) exists between 6 and 8 GPa before the full establishment of long range ordering above 8 GPa (see figure 25). Previous transport measurements (77) and recent new sets of experiment on high quality crystals (74) show that a low temperature Fermi liquid regime will appear below $T_A$ at a temperature roughly two order of magnitude smaller than the temperature $T_{MAX}$ of the resistivity maxima. Again, $T_A$ remains finite on entering in the magnetic quantum singularity; the observation of a pressure window for the appearance of magnetism supports strongly a M-FOT associated with a valence transition.

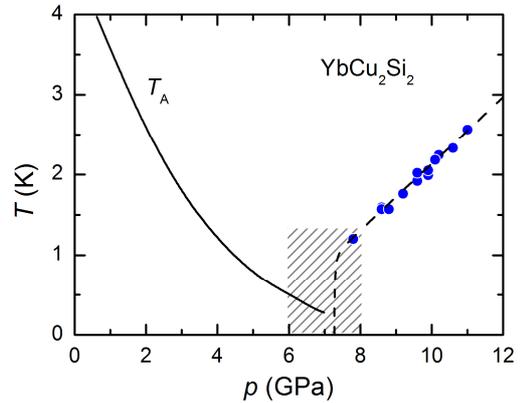

Fig. 25 : (T, P) phase diagram of YbCu$_2$Si$_2$. $T_A$ is the temperature below which a $AT^2$ resistivity law is observed. Let us notice again the finite value of $T_A$ in the pressure window (hatched area) where long range magnetic ordering occurs. As observed below for YbRh$_2$Si$_2$, a large temperature window will occur before the recovery of Fermi liquid properties (74,77).

**YbRh$_2$Si$_2$ : an exotic example, a puzzle difficult to resolve.**

In YbRh$_2$Si$_2$, the basal plane is an easy magnetization plane (78). The efficiency of the CF is clear from the high value of the $\chi_\perp / \chi_\parallel$ ratio near 30. The proposed CF scheme is four doublets located precisely at 0 , 17 , 25 and 43 meV while the Kondo temperature is estimated to $T_K$ = 25 K (79). Despite this high value, the surprising result is that YbRh$_2$Si$_2$ becomes long range magnetically ordered below $T_N$ = 70 mK (78, 80). Furthermore in face of the low residual entropy involved in the AF phase transition (S= 0.03 Rln2) the specific heat anomaly is very sharp (figure 26). Such a feature has never been observed in any Ce HFC near its M-QCP and also is quite different from molecular field behaviour suspected to work at very low temperature for itinerant magnetism. So, the onset of magnetism seems to point out another class of effects as if quasi free $Yb^{3+}$ centres survive with an associated large effective mass. The residual γ term is huge γ = 1.8 J mole$^{-1}$K$^{-2}$. The simple image is that even if there are few acting magnetic centres on a slow time window, the magnetic coupling is possible due to their slow velocity by comparison to the fast information transmitted by the light electrons. Another unusual observation is the detection of an electron spin resonance signal (81) obviously not given by the local dynamical response of the Yb (22).

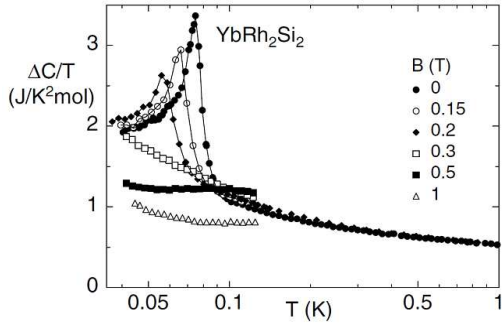

Fig. 26: At $H = 0$, a sharp specific heat anomaly of YbRh$_2$Si$_2$ at its ordering temperature $T_N$ ~ 70 mK with the remaining of a large γ ~ 1.8 Jmole$^{-1}$K$^{-2}$ term appears. The strong field variation of the specific heat is the mark of the proximity to quantum singularities (80).

Up to now attention was weakly given on the AF phase transition but was quickly turned on the field response (see the $H, T$ phase diagram figure 27). A field induced quantum criticality (80) was pointed out from the huge field dependence of $A(H)$ coefficient of the $T^2$ term of the resistivity which at zero order is directly proportional to $m^{*2}$. It has been claimed that $A(H)$ just above $H_C$ ~ 68 mT (along the easy magnetization plane of this tetragonal crystal) diverges at $H_C$ as it can be represented by the simple formula $1/(H-H_C)$. This assortment was refuted (22) in our own measurements as a maximum was found in the field variation of $A(H)$ at $H_C$ as represented figure (27). Like for the two previous cases of YbAuCu$_4$ and YbCu$_2$Si$_2$ under pressure, here $T_A(H_C)$ does not collapse again at $T \to 0K$ but intercepts the $T_N (H)$ line at a finite temperature.

Support on a abrupt change from a small to a large Fermi surface and thus on the validity of the local criticality approach was given from Hall measurements (82), but it was stressed that in this multiband material the link between a change in the Hall effect and drastic modification of the Fermi surface is not obvious (83). Here the real difficulty is the weakness of $H_C$ which prevents any hope to succeed in the full determination of the Fermi surface by quantum oscillations. Few quantum oscillations have been detected unfortunately in magnetic field far higher than $H_C$ (22, 84).

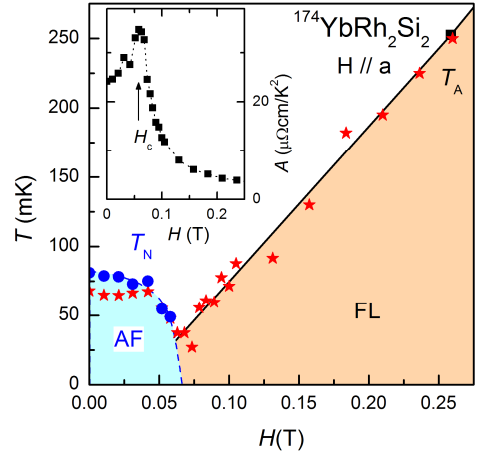

Fig. 27: ($H, T$) phase diagram of YbRh$_2$Si$_2$ with again a finite value of $T_A$ at the AF-PM boundary. Insert shows the $H$ evolution of the $A$ coefficient of the $AT^2$ resistivity term (22).

Recently extensive studies (85-88) have been continued in Dresden on the quantum criticality of YbRh$_2$Si$_2$ under pressure and magnetic field with precise data on the temperature dependence of the specific heat, the thermal expansion and the magnetization. The collection of systematic studies with accurate temperature dependence is useful for the comparison between theory and experiments. Our own feeling was rapidly that, in YbRh$_2$Si$_2$, the magnetic instability cannot be disconnected from the valence instability. This statement was already argued even for Ce HFC since it was observed that in the majority of examples the M-QCP at $P_C$ corresponds to the lost of the crystal field sensibility as at $P_C$, the Kondo temperature $k_BT_K$ is very often comparable to the CF splitting. To our knowledge there are only a few cases where at $P_C$, $k_BT_K$ remains smaller than $C_{CF}$ i.e. where only a spin $S = ½$ case is preserved through $P_C$. In the CeRu$_2$Si$_2$ and CeCu$_2$Si$_2$ families, fortunately $P_v - P_C$ does not vanish and corresponds to a 3-4 GPa window. Thus let us stress that the naive Doniach picture often used as a rigid dogma is in many cases irrelevant ("the Doniach syndrome" according to Prof. J. Akimitsu). For Ce HFC the interplay between magnetic and valence instabilities are often strong ($P_c \approx P_{CF}^* \approx P_v$). Again the novelty in the mixed valence phase of Yb compounds is that the crystal field strength may be preserved even for noticeable departure of $n_f$ from unity since now $T_K$ remains low enough.

A sound theoretical basis of the interplay between magnetic and valence transition in YbRh$_2$Si$_2$ was recently given by the extension of the previous model developed for field variation of the valence transition to a crossover regime (89). On this simple basis, the authors reproduce the magnetization variation through $H_C$ as well as different properties (fig. 28). The fancy properties of YbRh$_2$Si$_2$ will be the interplay of $T_N(H)$ and the crossover temperature $T_v(H)$ border directly linked with proximity of V-QCP.

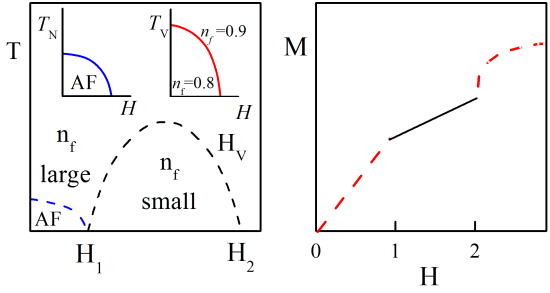

Fig. 28: $T_v$, $H_v$ crossover temperature and valence field (left) and the corresponding field variation of the magnetization (right) with a change of regime on crossing $H_1$ and $H_2$ (drawn from idea in (89)). Inserts (left panel) show the field dependence of $T_N$ and $T_v$ assuming bare phenomena, either magnetic or valence fluctuating.

As mentioned before by comparison to Ce HFC, the $P$ dependence of $T_N$ in YbRh$_2$Si$_2$ is quite anomalous (figure 15); the initial pressure increase of $T_N$ is followed by an intermediate $P$ regime centred around a shallow maximum at $P = 4.5$ GPa while a regime rather similar to the Doniach picture will occur only above $P^* = 9$ GPa, the pressure where basically the 3+ regime is really approached (22). The unusual pressure dependence of $T_N$ reflects the relative change of $U_{fc}$ by comparison to the other ingredients; this was described in ref. 22 via a change of the Kondo temperature of the 5d electron of Rh. Furthermore, the discontinuity observed in magnetoresistivity at $H = 2T$ for $P = 2.4$ GPa was recently taken of the mark to hit a V-CEP through the field scan (90).

To achieve a definite experimental conclusion on the YbRh$_2$Si$_2$ case remains a risky goal due to the weakness of $T_N$ and of $H_C$. As for Ce HFC with the emergence of the 115-family, the main progress will arrive with a discovery of a clear case where the Fermi surface can be fully determined. The best example will be an Ising type case with positive value of $P_C$ and a high value of $H_C$ governed by its finite magnetic field critical end point. On theoretical side, the example of YbRh$_2$Si$_2$ has pushed to new perspectives on quantum singularities including CEP, TCP and FOQT, i.e leads to escape to the naïve view of second order quantum critical points (90).

## β-YbAlB$_4$: the first Yb based heavy fermion superconductor

The evidence of strong valence fluctuation in β-YbAlB$_4$ was recently found in hard X ray photoemission study with the Yb valence estimated around 2.74 (91). The interest in this system comes from the first discovery of superconductivity in an Yb based heavy fermion compound (92). This observation appeared only recently despite searching over two decades to find a hole SC analogue of the Ce examples. In zero magnetic field, up to a crossover temperature near 2 K, the specific heat show a $\log T$ non-Fermi liquid increase on cooling down to 80 mK where SC appears (figure 29); the residual γ term of the normal phase is near 200 mJ mole$^{-1}$K$^{-2}$; no $T^2$ dependence of the resistivity is observed down to $T_C$ but a $T^{3/2}$ variation which may be characteristic of the proximity form AF-QCP. Furthermore applying a magnetic field changes again the shape of the specific heat at low temperature as well as the power of the $T^n$ dependence of the resistivity.

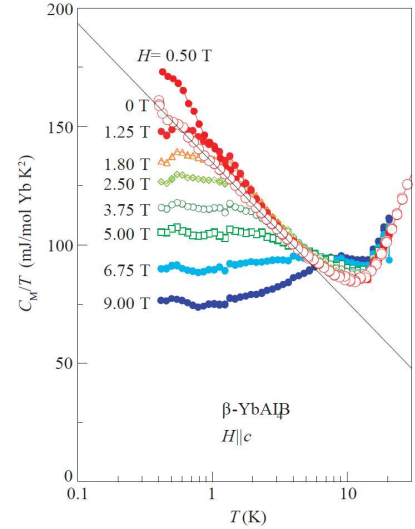

Fig. 29: Non Fermi liquid behaviour of the normal phase in the first Yb based HFC superconductor β-YbAlB$_4$ (92).

Recent measurements of the upper critical field (93) give for T → 0K an extrapolation for field in the a, b plane $H_{c2} \approx 150$ mT and along the c axis near $H_{c2} \approx 25$ mT (figure 30). From the initial slope of H$_{c2}$ at $T_C$, the respective orbital limits are 150 mT and 76 mT while the Pauli limit due to the Zeeman effect on the Cooper pair predicts $H_{C2}^P = 150$ mT i.e six time higher than the extrapolated value of 25 mT along the easy c magnetization axis. The huge enhancement in the paramagnetic limitation deserves further explanation. It is interesting to note that here again as for the previous Yb cases (YbInCu$_4$, YbCu$_2$Si$_2$ and YbRh$_2$Si$_2$) ferromagnetic enhancement in the low field susceptibility is detected at low temperature in agreement with the prediction given near V-QCP. As noted by the authors, it is in contrast with the cases of many Ce based compound however this FM enhancement does not preclude the proximity of AF which may driven by another source of intersite couplings.

Concerning SC, the discovery of SC in β-YbAlB$_4$ breaks the feeling that the observation of unconventional SC in Yb HFC will be difficult (as searching for a needle in a bundle of hay)... Outside SC, β-YbAlB$_4$ is also interesting as at $P = 0$ it is assumed to be in a paramagnetic ground state (figure 29) with non-Fermi liquid behaviour). Under pressure, long range magnetism must appear and was recently reported (94) at $P_c$ around 3 GPa with a rather high value of the magnetic ordering temperature (25K). Thus it is another case to study the interplay between SC, valence and magnetic fluctuations. However the low values of $T_c$ and of the Fermi liquid temperature which are already below 80 mK at $H = 0$ at $P = 0$ point out the difficulty of the task. The rush to a better example has already started. Here again the hope is to discover a SC material with high enough $T_c$ for deep studies on their SC properties. A further goal is to find also a material where large crystal can be realized in order to attack the problem with multiple approaches.

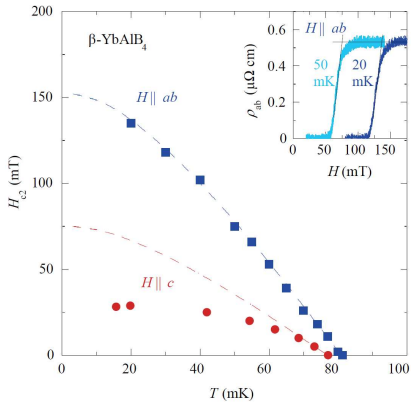

Fig. 30: Observation of superconductivity in β-YbAlB$_4$ for different magnetic fields. The temperature dependence of the upper critical field $H_{c2}$ along the ab plane (square, blue) and the c axis (circle, red). Different single crystals with similar $T_c$ were used for each field direction measurement. Broken lines represent the curves obtained by fitting to the experimental results near $T_c$ using the theoretical model. Inset: In-plane resistivity vs $H$. $H_{c2}$ is defined as the crossing point of the linear extrapolation lines of the resistivity in the normal state and the resistivity change in the transition (94).

## 10/ Conclusion

In this article, we have tried to present the power of HFC as a material to clarify the field of quantum phase transitions. Ce HFC and also U HFC give already a sound basis for deep studies in order to understand microscopic mechanism as well as new macroscopic effects. The opening to the family of Yb HFC may lead to new phenomena as the sharpness of the level (i.e Δ) will give the opportunity to clarify major problems such as debates on global or local quantum magnetic singularities, on the interplay between magnetic, superconducting and valence instabilities. For Yb HFC, the great chance to preserve an operational crystal field effect even for relatively small $n_f$ may be a crucial point. Up to now this important issue has never been considered deeply theoretically.

We hope to have convinced the reader that precise experiments on extreme conditions (very low temperature, high magnetic field and high pressure) are now possible with qualities approaching that of date obtained on quantum liquid and solid He$^3$. Finally it is worthwhile to underline that progresses in the experimental accuracy have required often breakthroughs in instrumentation. There were initially initiated by a physical target but they are now available to quite different fields even outside the condensed matter community.

Finally let us stress again that HFC results are sound basis in the large domain of strongly correlated electronic systems but also in domains going from cosmology (95) to cold fermionic gases (96-97).

### Acknowledgements
JF thanks K Miyake for many comments, my Japanese colleagues from Osaka, Kyoto, and Tokyo universities for stimulating discussions (K Kitaoka , Y Onuki, K Ishida, K Izawa, H Sato, S Watanabe) as well as my colleagues in Grenoble (D Aoki, G Knebel, T Matsuda, JP Sanchez). He uses this opportunity to express his gratitude to Pr K Asayama, Pr T Kasuya, Pr Y Miyako and Pr J Sakurai for opening the gate to Japan more than two decades ago.